\documentstyle[preprint,epsf,aps,prl]{revtex}
\begin{document}

\title{Annular electroconvection with shear}
\author{Zahir A. Daya,$^{1}$ V.B. Deyirmenjian,$^{1}$ Stephen W. Morris,$^{1}$
and John R. de Bruyn.$^{2}$}
\address{$^{1}$Department of Physics and Erindale College, University of
Toronto, 60 St. George St., Toronto, Ontario, Canada M5S 1A7\\
$^{2}$Department of Physics and Physical Oceanography, Memorial University of
Newfoundland, St. John's, Newfoundland, Canada A1B 3X7}
\date{\today}
\maketitle

\begin{abstract}
We report experiments on convection driven by a radial electrical force in
suspended annular smectic A liquid crystal films. In the absence of
an externally imposed azimuthal shear, a stationary one-dimensional (1D)
pattern consisting of symmetric vortex pairs is formed via a supercritical
transition at the onset of convection. Shearing reduces the symmetries of the
base state and produces a traveling 1D pattern whose basic periodic unit is a
pair of asymmetric vortices. For a sufficiently large shear, the primary
bifurcation changes from supercritical to subcritical. We describe measurements
of the resulting hysteresis as a function of the shear at radius ratio $\eta
\sim 0.8$. This simple pattern forming system has an unusual combination of
symmetries and control parameters and should be amenable to quantitative
theoretical analysis.
\end{abstract}

Extended nonlinear dissipative systems can develop complicated spatial and
temporal patterns when subjected to external stresses by the variation of
a control parameter\cite{ch93}. The observed patterns are often the
result of symmetry-breaking bifurcations. In some cases, additional
control parameters can alter the symmetries of the base state which in
turn affects the bifurcation to the patterned state.
For example, in a large cylinder, the onset of Rayleigh-B\'{e}nard
convection (RBC) is stationary. However, rotating the cylinder breaks the
reflection symmetry in any vertical plane containing the rotation axis and
convection occurs via a Hopf bifurcation\cite{rotRBC}. Varying the rotation
rate can move the system into a regime of spatio-temporal
chaos\cite{hs96}. It is useful to identify other simple situations in which
pattern forming bifurcations can be ``tuned" by additional control
parameters in order to better understand how we can manipulate spatiotemporal
patterns.

In this Letter, we exploit the novel properties of electroconvection in freely
suspended smectic~A liquid crystal films in an annular geometry to study how
azimuthal shear affects pattern formation. This is accomplished by measuring
the current-voltage characteristics of films subjected to a dc voltage
between the inner and outer edges of the annulus. A variable Couette shear is
applied by rotating the inner edge of the annulus. The films had a radius
ratio $\eta = r_i / r_o \sim 0.8$, where $r_i$ ($r_o$) is the inner (outer)
radius. For zero shear, a stationary one-dimensional (1D) pattern, consisting
of pairs of symmetric vortices, appears via a continuous bifurcation at a
voltage $V_c^o$. For non-zero shear, an azimuthally traveling 1D pattern, whose
basic unit is a pair of unequally sized vortices, is established. When
the shear is of sufficient magnitude, the primary bifurcation becomes
discontinuous. The combination of applied shear, radial electrical
driving forces, and a two-dimensional (2D) annular geometry makes it possible
to adjust the symmetries of the base state and the nature of the primary
bifurcation.

Electroconvection in suspended smectic films without applied shears has been
the subject of several previous experimental and theoretical
studies\cite{90prl,jstatphys,91pra,endsel,cpip97,linstab97,gle97,smctheory,smc}.
Smectic liquid crystals are
materials which consist of layers of orientationally ordered molecules. In
smectic A, the long axis of the molecules is normal to the layer plane. In this
arrangement, smectic A exhibits 2D isotropic fluid properties in the layer
plane and flow normal to the layers is strongly
suppressed\cite{90prl,jstatphys,91pra,endsel,cpip97}.
Finite rectangular smectic A
films were the subject of previous experimental
work\cite{90prl,jstatphys,91pra,endsel,cpip97}, while
the linear and weakly nonlinear theory have been completed for the case of
a 2D weakly conducting fluid in a laterally unbounded rectangular
geometry\cite{linstab97,gle97}. Convection has also been studied in smectic
C films in which the molecular axes are tilted with respect to the layer
normal and can be reoriented by the flow\cite{smctheory,smc}. These
anisotropic effects will not concern us here.

There are several features of this system which distinguish it from the
well-studied cases of RBC and Taylor Vortex Flow (TVF).  The 2D nature of the
flow severely restricts the secondary instabilities available to the
pattern\cite{endsel}. A Couette shear flow cannot easily be superposed on
RBC as is accomplished in annular electroconvection. This is also one of
a handful of pattern forming systems in
which the dynamics are governed by interactions mediated by long range forces.
Convection is driven by the interaction of the self-consistent electric field
with a non-zero surface charge density on the film\cite{linstab97,gle97}.

An additional consequence of the shear is the imposition of an azimuthal mean
flow. RBC with horizontal mean flow and TVF with axial mean flow have been
investigated experimentally and theoretically\cite{ch93,mt95,bac94}. These
open flow studies showed that the first instability is {\em
convective}\cite{ch93} and that the pattern drifts in
the direction of the flow. In contrast, our annular
system is closed so that the first instability is {\em absolute}. It is
interesting to note that, unlike in 3D TVF, an imposed Couette shear flow
alone does not lead to an instability in a 2D fluid\cite{wmkg95}. We show
below that shears always have the effect of suppressing electroconvection.

This system also possesses interesting symmetries.
In the absence of shear, the base state is invariant under azimuthal rotation
and reflection in any vertical plane containing the rotation axis through the
center of the annulus. The application of azimuthal shear breaks the
latter reflection symmetry and distinguishes between the clockwise and
counter-clockwise directions. This results in a pattern at the
onset of electroconvection which travels in the direction of the mean flow.

Our experimental apparatus consisted of two stainless steel electrodes.  The
inner electrode was a circular disk of diameter $8.96\pm0.01$~mm.  The outer
electrode was a larger circular plate with a central hole of diameter $11.20
\pm 0.01$~mm.  The annular film was suspended between the concentric
inner and outer electrodes, giving a radius ratio $\eta = 0.800 \pm 0.001$. A
motorized stainless steel blade wetted with liquid crystal was slowly drawn
across the annulus to form a film.  This procedure allowed for the drawing of
uniform films of various thicknesses. The films were composed of integer
numbers of smectic layers.  Each layer of smectic A 8CB\cite{8cb} is $3.16$~nm
thick. The experiments employed films of up to~75 layers. The films were viewed
through a microscope with a color video camera.  When viewed in reflected white
light, the films displayed a color due to interference. By visually monitoring
the film color throughout each experiment, we confirmed that the films had a
constant thickness over their whole area which remained uniform to within $\pm
1$ layer.  The inner electrode was rotated about its axis by a high
precision stepper motor at angular frequencies up to
$\omega \sim 10$~rad/s.  All experiments
were performed at $23 \pm 1^\circ$C, well below the smectic
A-nematic transition at $33^\circ$C for pure 8CB.  The system was shielded
by a Faraday cage and the outer electrode and shield were maintained at ground
potential while the voltage of the inner electrode was varied above ground. The
resulting current through the film was measured using a picoammeter.

\begin{figure}
\epsfxsize=4in
\centerline{\epsffile{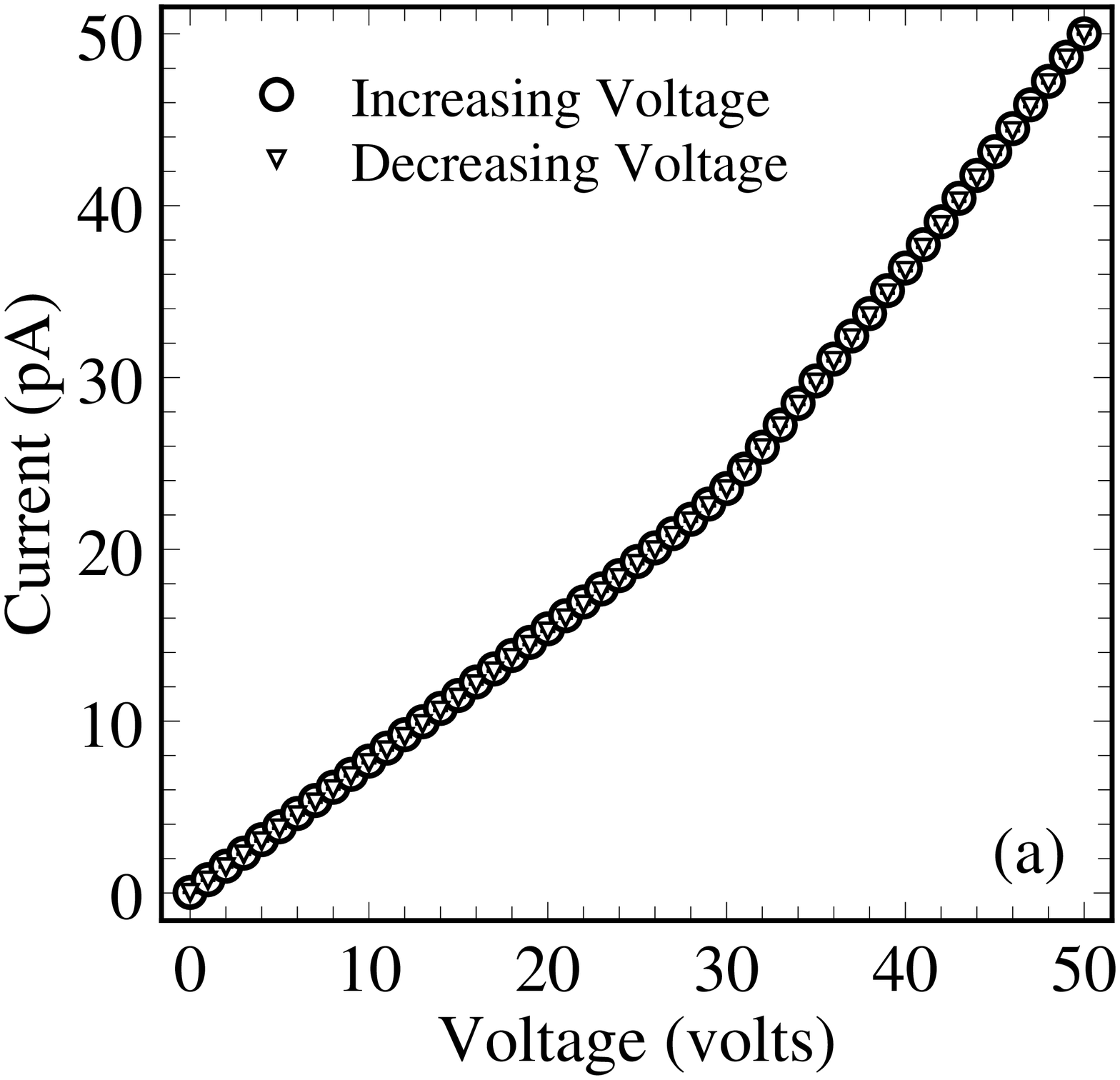}}
\epsfxsize=4in
\centerline{\epsffile{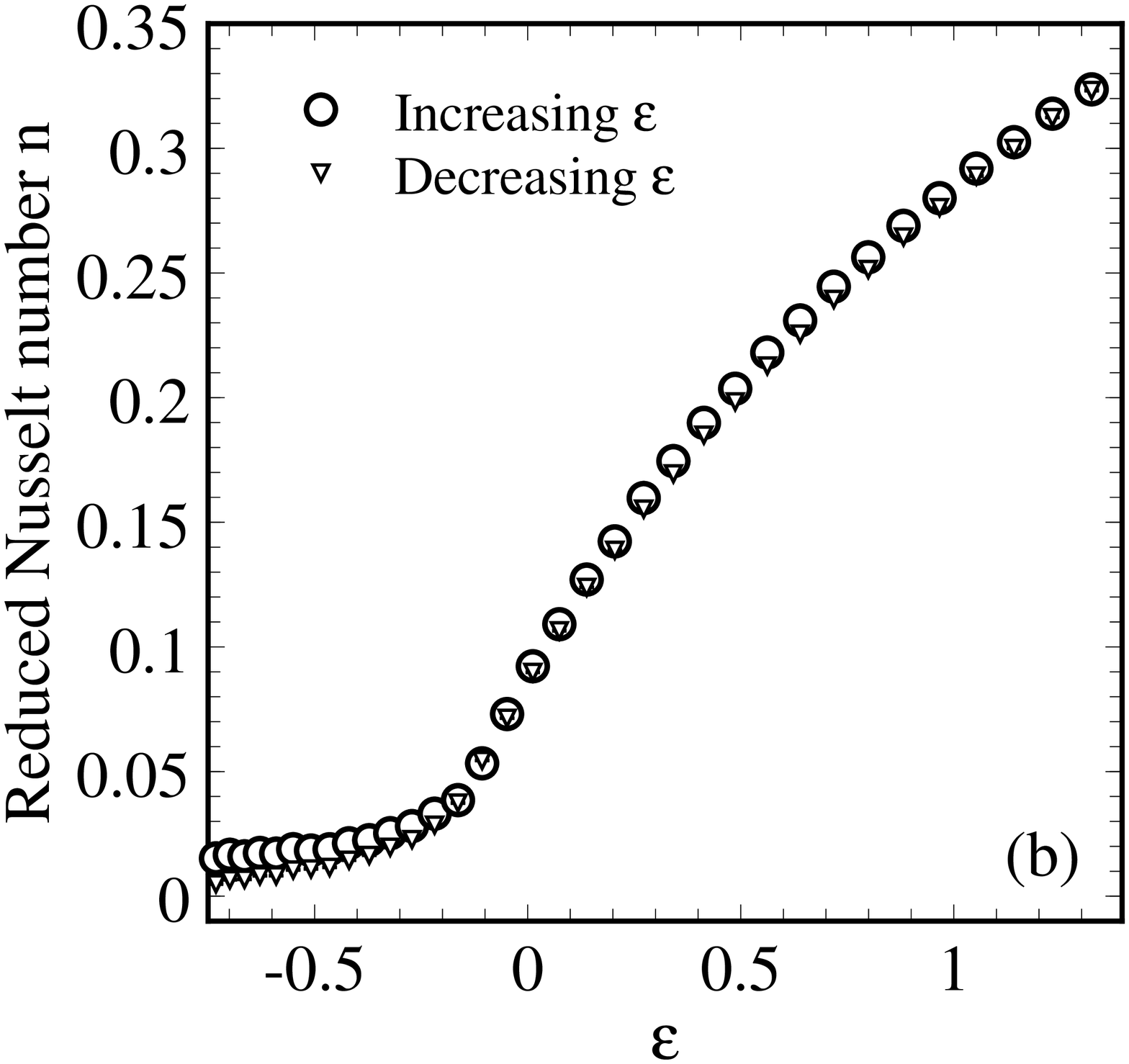}}
\vskip 0.1in
\caption{(a) The current-voltage measurements for an annular film with
radius ratio $\eta \sim 0.8$. (b) The reduced Nusselt number $n$, versus the control parameter $\epsilon$.}
\label{IV}
\end{figure}

A plot of a typical current-voltage response in the absence of shear is
shown in Fig.~\ref{IV}(a). These data were obtained by slowly increasing and
then decreasing the applied voltage in small steps. After each voltage
step, the film was allowed to relax for about 10 seconds, before 20-25
measurements of the current,
each separated by 250 milliseconds, were averaged. The film behaved as an
ohmic conductor in the conduction regime below the onset of electroconvection.
At a critical voltage $V_c^o$, the film was unstable to a stationary vortex
pattern which broke the continuous axisymmetry of the base state.  The fluid
flow could be visualized by allowing fine dust to settle on the film. The
pattern was comprised of pairs of equally sized, symmetric,
counter-rotating vortices. $V_c^o$ was identified by the position of the
kink in the current-voltage response (see Fig.~\ref{IV}(a)). No
hysteresis in the onset was observed between runs with increasing and
decreasing voltage. These data are similar to those found previously for
rectangular films, but are much more precise, since annular films do not
suffer from the problem of leakage of current around the ends of the
film\cite{91pra}.

In analogy with the convective heat transport in RBC, we describe the
dimensionless convective current by a reduced Nusselt number $n = I/I_c -
1$, where $I$ ($I_c$) is the total (conductive) current. The
dimensionless control parameter for this system is\cite{linstab97,gle97}
${\cal R} = \epsilon_0^2
V^2/\sigma \mu s^2$, where $\epsilon_0$ is the dielectric constant of free
space, $\sigma$ is the bulk conductivity, $\mu$ is the bulk molecular
viscosity, and $s$ is the film thickness. We employ the reduced
control parameter $\epsilon = {\cal R}/{\cal R}_c^o -1 = (V/V_c^o)^2 - 1$. When
recast in terms of $n$ and $\epsilon$, the current-voltage data takes the form
shown in Fig.~\ref{IV}(b), which is typical of a supercritical
bifurcation\cite{drift}.

The number of vortex pairs that is generated in
the unsheared annulus equals the number that would appear in a rectangular film
whose length equals the circumference of the annulus\cite{jstatphys}.  We
expect, however, the range of stable vortex wavelengths in the annulus will
be different from the rectangular case.  In the latter, the range is
determined by the boundary conditions at the ends of the film\cite{endsel}.
In the annulus, the 1D pattern is effectively subject to periodic boundary
conditions.

\begin{figure}
\epsfxsize=4in
\centerline{\epsffile{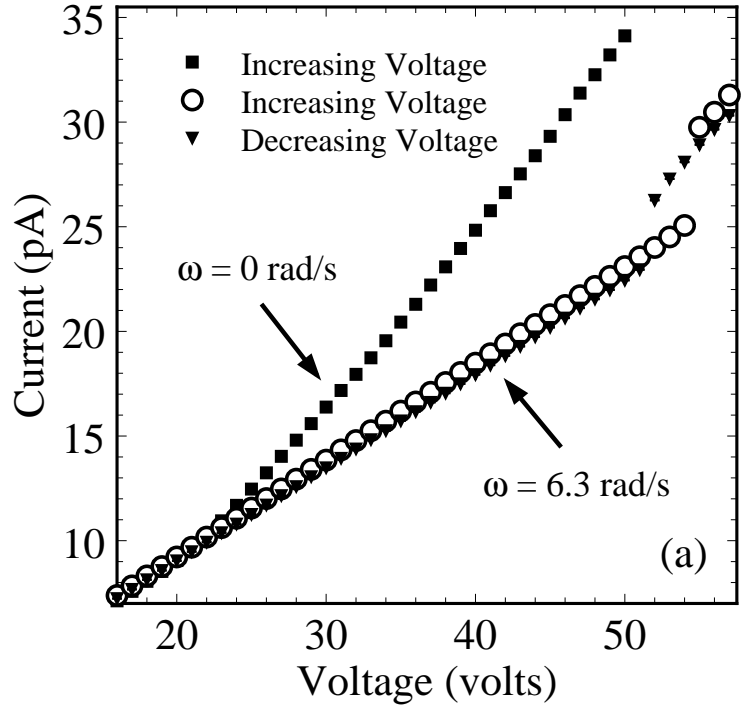}}
\epsfxsize=4in
\centerline{\epsffile{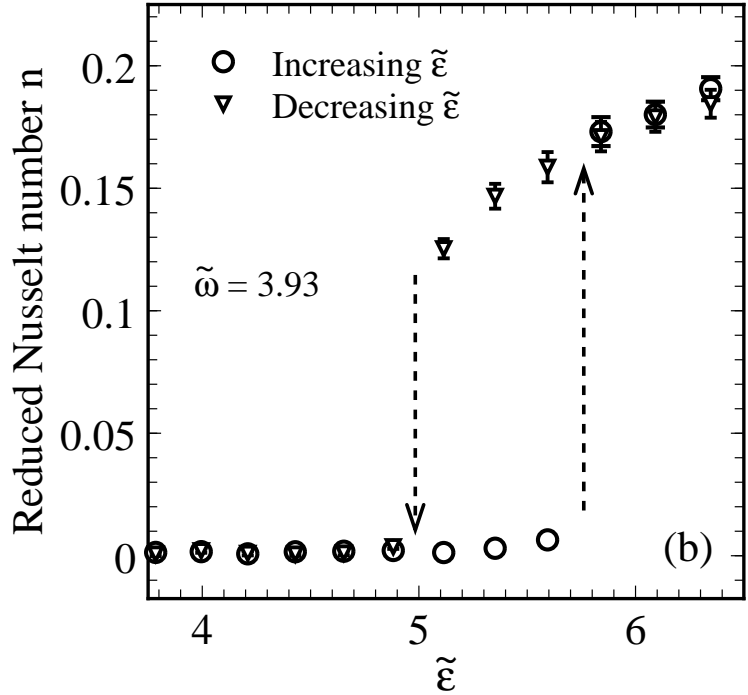}}
\vskip 0.1in
\caption{(a) The current-voltage measurements for the same film with and without applied shear. (b) The reduced Nusselt number $n$ vs.
$\tilde{\epsilon}$ for a film sheared at dimensionless angular frequency $\tilde{\omega}=3.93$. }
\label{NEW}
\end{figure}

When the inner electrode is rotated about its axis, the film is subjected to a
steady shear. The sheared film was allowed about 30 seconds to attain a steady
state, after which current-voltage data were acquired as described earlier.
Figure~\ref{NEW}(a) compares the current-voltage characteristic for the
same film when the base state is quiescent or sheared. For small shears,
the onset of convection occurs by a continuous transition and is identified by
a kink in the current-voltage data, as in the zero shear case. When the shear
is sufficiently large, the transition is subcritical and discontinuous and is
revealed by a jump in the current. In this regime, the bifurcation is
hysteretic with convection appearing at $V_c^+(\omega)$ as $V$ is
increased, and disappearing at $V_c^-(\omega)$ as $V$ is decreased.
The onset of convection is suppressed by the shear as
$V_c^{\pm}(\omega) > V_c^o$.

\begin{figure}
\epsfxsize=4in
\centerline{\epsffile{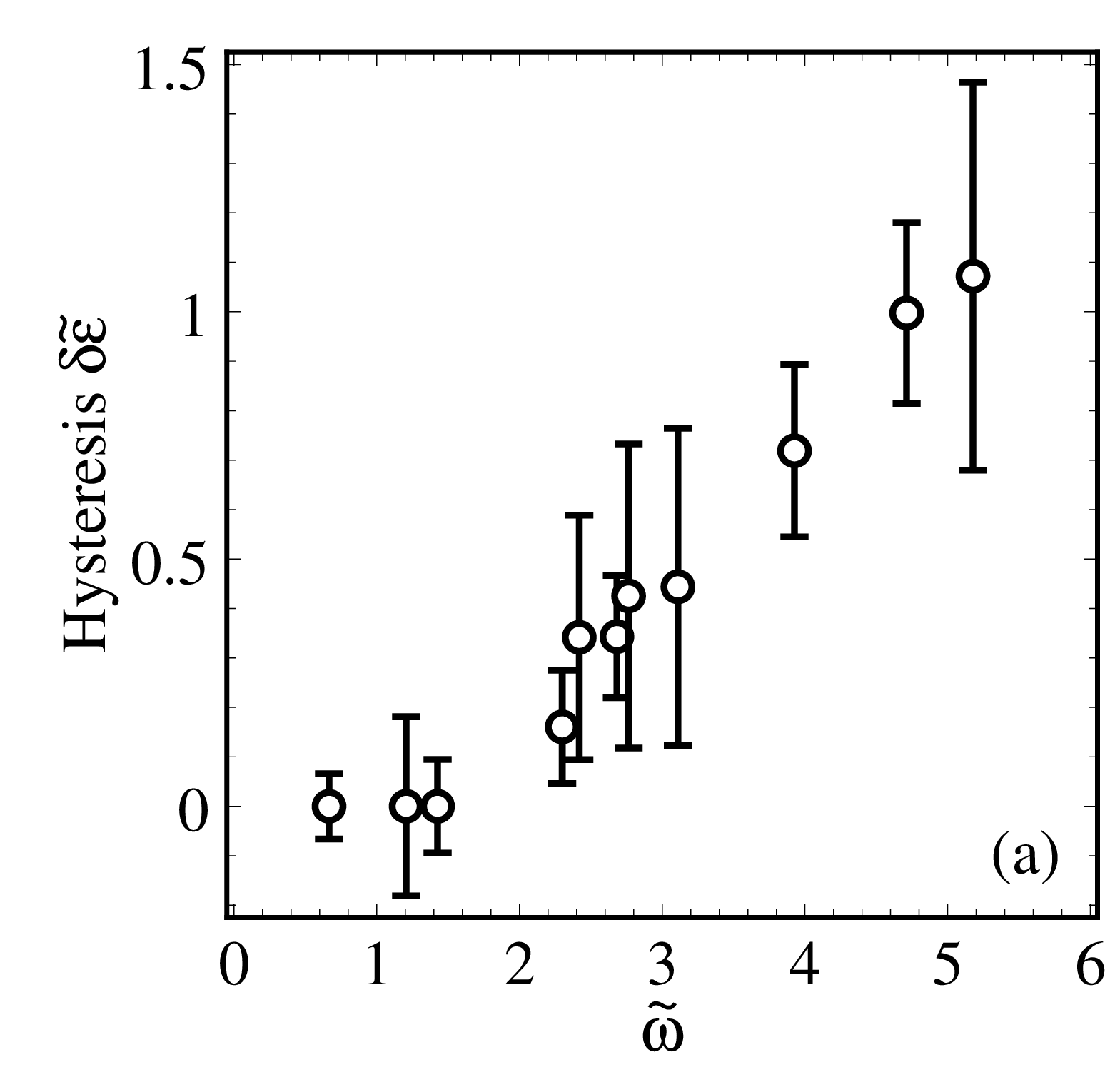}}
\epsfxsize=4in
\centerline{\epsffile{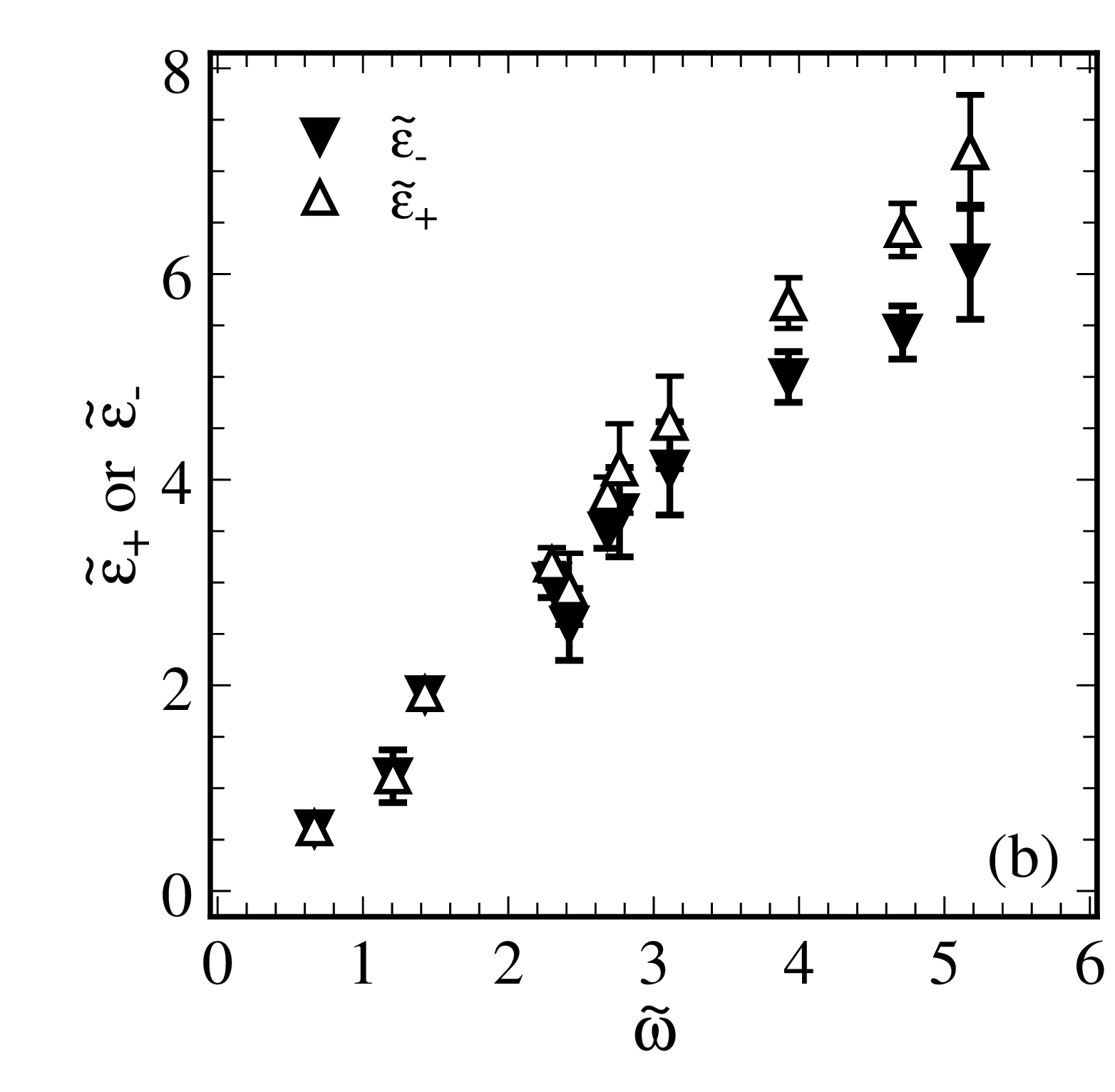}}
\vskip 0.1in
\caption{(a) Hysteresis $\delta \tilde{\epsilon}$ versus the dimensionless angular frequency $\tilde{\omega} $. (b) Measurements of
$\tilde{\epsilon}_+$ and $\tilde{\epsilon}_-$ vs. $\tilde{\omega}$.}
\label{HYS}
\end{figure}

Figure \ref{NEW}(b) is a plot of $n$ versus
$\tilde{\epsilon}={\cal{R}}(\omega)/{\cal{R}}_c^o-1=(V(\omega) /V_c^o)^2 -1$
for a sheared film. The transition from the conducting (convecting) state to
the convecting
(conducting) state occurs at $\tilde{\epsilon}_+$ ($\tilde{\epsilon}_-$), where
$\tilde{\epsilon}_{\pm} = (V_c^{\pm}(\omega) /V_c^o)^2 -1$. The hysteresis,
${\delta \tilde{\epsilon} = \tilde{\epsilon}_+ - \tilde{\epsilon}_-}$, is
shear dependent.  We non-dimensionalize the rotation rate by the charge
relaxation time characteristic of the annular
film\cite{91pra,linstab97,gle97}, $\tau =
\epsilon_0 (r_o - r_i)/\sigma s$.  From the current-voltage
characteristic of the film in the conduction regime, the product $\sigma s =
\ln(r_o/r_i)/2\pi\rho$, where $\rho$ is the film resistance.  We therefore
define the dimensionless angular frequency as
$\tilde{\omega} = 2\pi \omega \epsilon_0 (r_o - r_i) \rho /\ln(r_o/r_i)$.
Our results for ${\delta \tilde{\epsilon}}$ in eleven experiments at various
$\tilde{\omega}$ are plotted in Fig.~\ref{HYS}(a). For
$\tilde{\omega} \lesssim 1.7$, the hysteresis is consistent with zero which
indicates that the bifurcation is continuous for small shears.  In
Fig.~\ref{HYS}(b) are shown measurements of $\tilde{\epsilon}_+$ and
$\tilde{\epsilon}_-$. We find that there is strong suppression of
convection by the shear, with $\tilde{\epsilon}_+ >$ 7 for the largest
$\tilde{\omega}$ studied.

The traveling pattern in the sheared experiment is most simply described in a
frame that rotates at the angular speed at which the pattern appears
stationary. In this frame it consists of a periodic array of vortex pairs.
Each pair is asymmetric, composed of counter-rotating vortices of unequal
width.
The vortex whose circulation is in the same direction as the
rotation of the inner electrode is narrower while the opposite sense vortex is
broader. The number of traveling vortices observed depends hysteretically on
$\tilde{\epsilon}$ and $\tilde{\omega}$ and is generally smaller than in the
unsheared case.  Transitions between different numbers of vortices can be
observed as $\tilde{\epsilon}$ and $\tilde{\omega}$ are varied.  The sheared
base state is invariant under azimuthal rotations, but not under reflections.
The onset of convection breaks the azimuthal symmetry and the resulting
traveling pattern has no continuous spatial symmetries.

In summary, we measured the current-voltage characteristics of radially driven
electroconvection in freely suspended annular smectic A films under varying
degrees of azimuthal shear. Shearing, which was generated by rotating the
inner edge of the annulus at an angular frequency $\tilde{\omega}$, altered the
symmetries of the system and changed the primary bifurcation from stationary,
when
$\tilde{\omega}=0$, to oscillatory. Adjusting $\tilde{\omega}$ allowed us to
tune the primary symmetry-breaking bifurcation and suppress the onset of
convection. Without shear, the onset of electroconvection was
supercritical, but the transition became subcritical with a large enough shear.
The pattern consisted of symmetric (asymmetric) vortex pairs when
$\tilde{\omega}=0$ ($\tilde{\omega} \neq 0)$. Future work on this system
will involve an experimental study of the secondary instabilities of the
pattern.  It is also interesting to explore the effect of smaller radius
ratios. This pattern forming instability should be simple enough to be
captured by a quantitative nonlinear theory.

We are grateful to T.C.A. Molteno, W.A. Tokaruk, K. Choo, W. Langford, and
M. Krupa for helpful discussions and assistance during the course of this work.
This research was supported by the Natural Sciences and Engineering Research
Council of Canada.

\vfill\eject

\vfill\eject

\vfill\eject


\begin{references}
\bibitem{8cb} To control the nature of the ionic species in the liquid crystal,
octylcyanobiphenyl (8CB), was doped with tetracyanoquinodimethane (TCNQ).
\end{references}
\end{document}